\documentclass[twocolumn,aps,superscriptaddress,prl,showpacs,floatfix]{revtex4}

\usepackage{graphicx,amsmath,amssymb}
\usepackage{bm}

\newcommand{\hh}{\hat{H}}
\newcommand{\hU}{\hat{U}}
\newcommand{\hDE}{\hat{\Delta E}}
\newcommand{\hrho}{\hat{\rho}_0}
\newcommand{\hUd}{\hat{U}^\dagger}
\newcommand{\delh}{\Delta\hat{H}}
\newcommand{\hA}{\hat{A}}
\newcommand{\hr}{\hat{\rho}}
\newcommand{\DF}{\Delta F}
\newcommand{\DE}{\Delta E}
\newcommand{\DJa}{\Delta J}

\newcommand{\om}{\omega}

\begin{document}

\title{Jarzynski equation for a simple quantum system: Comparing two
  definitions of work}

\author{A. Engel}
\email{engel@theorie.physik.uni-oldenburg.de} 
\author{R. Nolte}

\affiliation{Institut f\"ur Physik, Carl-von-Ossietzky-Universtit\"at,
     26111 Oldenburg, Germany }


\begin{abstract}
  The validity of the Jarzynski equation for a very simple,
  exactly solvable quantum system is analyzed. The
  implications of two different definitions of work proposed in the
  literature are investigated. The first one derives from measurements
  of the system energy at the beginning and at the end of the process
  under consideration making the work a classical stochastic variable
  with transition probabilities derived from quantum mechanics. In the
  second definition an operator of work is introduced and the average
  in the Jarzynski equation is a quantum expectation value. For the
  first definition a general quantum mechanical version of the
  Jarzynski equation is known to hold. For the second one the
  Jarzynski equation fails to yield the free energy difference at low
  temperature.
\end{abstract}

\maketitle

\section{Introduction}
The fluctuation properties of physical systems driven strongly away
from thermal equilibrium are of central importance in statistical
mechanics. Recent progress in this field gathers from a handful of
exact and remarkably general theorems characterizing the properties 
of exponentially rare trajectories. Although closely related to each
other it became customary to divide them into two groups:
non-equilibrium work theorems relating the work done in an
irreversible process to the free-energy difference between initial 
and final state \cite{BoKu,Jar,Jar2,HaSa}, and non-equilibrium
fluctuation theorems concerning the ratio between the
probabilities of entropy-producing and entropy-consuming processes 
\cite{EvCoMo,GaCo,Kurchan,JarFT,Crooks,Seifert,CBK}. 

Several proofs of these theorems are by now available if the
microscopic dynamics is described in terms of classical
mechanics. Particular examples include micro-canonical Hamiltonian
dynamics \cite{GaCo,CBK}, Hamiltonian dynamics coupled to different types
of thermostats \cite{GaCo,JarFT}, and stochastic dynamics of Langevin 
type \cite{Kurchan,Crooks,Seifert}.  

The situation is less satisfactory for the case in which the system
is described by quantum mechanics at the microscopic level 
\cite{Yukawa,Kurchan1,Tasaki,Mukamel,Mukamel2,EsMu,Maes,JaWo,RoMa,
MoTa,Monnai,AlNi}. Central problems in generalizing non-equlibrium
theorems of the described 
type to the quantum realm are the definition of the work performed on
the system in the irreversible transition and the identification of
the proper average to be taken. If the average is interpreted as an 
average over quantum mechanical transition probabilities for the
evolution between initial and final Hamiltonian 
\cite{Kurchan1,Tasaki,Mukamel,EsMu,Mukamel2} the description of the quantum
mechanical system is formally similar to a classical one and
quantum versions of the fluctuation theorems can be proven
in analogy to the classical case. Alternatively the work may be
related to the difference of the Hamiltonian of the system
in the Heisenberg representation at different times
\cite{BoKu,Yukawa,AlNi,AlNi2}. The average to be 
performed is then a quantum mechanical expectation value with the
initial density operator.

In the present note we investigate the validity of a limiting case of
the Jarzynski equation for a very simple quantum system. This system
is exactly solvable and physically transparent and the implications of
both definitions of work and corresponding averages can be studied 
analytically. We will show that both approaches are compatible with
the second law of thermodynamics and correctly describe the
classical limit $\beta \hbar\to 0$ but yield different results in the
quantum regime. We will indicate the formal as well as the physical
reason for the failure of the Jarzynski equation in the second
approach which we nevertheless believe to be the more appropriate 
one.

\section{The model}
We consider a one-dimensional quantum harmonic oscillator of mass $m$ and use
units in which $m=\hbar=1$. The Hamiltonian is given by 
\begin{equation}
  \label{eq:defH}
  \hh_0=\frac{\hat{p}^2}{2} + \frac{\om_0^2}{2}\,\hat{x}^2
\end{equation}
with the usual meaning of the symbols. The system is supposed to be at
equilibrium with a heat bath of inverse temperature $\beta$. From
elementary quantum statistics the free energy is given by 
\begin{equation}
  \label{eq:F0}
  F_0=\frac{\om_0}{2}+\frac{1}{\beta}\ln(1-e^{-\beta\om_0})\; .
\end{equation}
The system is now decoupled from the heat bath and the frequency of
the oscillator is changed from $\om_0$ to $\om_1$ according to 
\begin{equation}
  \label{eq:defom}
  \om^2(t)=\om^2_0+(\om^2_1-\om^2_0)\frac{t}{t_f}\, ,
\end{equation}
where $t_f$ denotes the final time of the process. For $0<t<t_f$ the
system is hence characterized by a unitary time evolution described by
\begin{equation}
  \label{eq:defU}
  \hU(t)= \mathrm{T}_> \exp\Big(-\frac{i}{\hbar}\int_0^t dt' \, \hh(t')\Big)
\end{equation}
induced by the time-dependent Hamiltonian
\begin{equation}
  \hh(t)=\frac{\hat{p}^2}{2} + \frac{\om^2(t)}{2}\,\hat{x}^2\; ,
\end{equation}
where $\mathrm{T}_>$ denotes the time ordered product. The
classical version of this system was discussed in \cite{Jar3,Jar4}. 

Since there is no coupling to the heat bath during the process the
work $W$ necessary to accomplish the change $\om_0 \mapsto
\om_1$ equals the change in energy $\DE$ of the system. The
Jarzynski equation \cite{Jar} hence stipulates  
\begin{equation}
  \label{eq:JE}
  \langle e^{-\beta\DE}\rangle=e^{-\beta\DF}\; ,
\end{equation}
where $\langle ...\rangle$ denotes the average over repeated
realizations of the process and 
\begin{equation}
  \label{eq:DF}
 \DF=\frac{1}{\beta}\ln\frac{\sinh\frac{\beta\om_1}{2}}
                            {\sinh\frac{\beta\om_0}{2}}
\end{equation}
as follows from (\ref{eq:F0}). 

\section{Measuring energy twice}
The definition of work advocated in
\cite{Kurchan1,Tasaki,Mukamel,Mukamel2,TLH} 
derives from a measurement of the energy of the system at the
beginning and at the end of the process. Measuring $\hh$ at $t=0$
yields an eigenvalue $E_n^{(0)}$ of $\hh_0$ with probability 
\begin{equation}
  \label{eq:defP0}
  P_n^{(0)}=e^{\beta (F_0-E_n^{(0)})}\; ,
\end{equation}
and leaves the system in the corresponding eigenstate
$|\psi_n^{(0)}\rangle$  of $\hh_0$. At time $t_f$ the system is hence
in state $\hU(t_f) |\psi_n^{(0)}\rangle$ and measuring $\hh$ anew
yields an eigenvalue $E_m^{(1)}$ of $\hh_1=\hh(t_f)$ with probability 
\begin{equation}
  \label{eq:defWmn}
  W_{mn}=|\langle \psi_m^{(1)}|\hU(t_f)|\psi_n^{(0)}\rangle|^2\; ,  
\end{equation}
whith $|\psi_m^{(1)}\rangle$ denoting the corresponding eigenstate of
$\hh_1$. The energy difference $\DE$ is now identified with 
$E_m^{(1)}-E_n^{(0)}$ and the average of a function $f(\DE)$ is
defined as
\begin{equation}
  \label{eq:clav}
  \langle f(\DE)\rangle=\sum_n P_n^{(0)} \sum_m W_{mn} \;
          f(E_m^{(1)}-E_n^{(0)})
\end{equation}
One then easily finds \cite{Kurchan1,Tasaki,Mukamel}  
\begin{align}\label{wrong1}
  \langle &e^{-\beta\DE}\rangle=\sum_n P_n^{(0)} \sum_m W_{mn} \; 
     e^{-\beta\DE}\\\nonumber
 &=e^{\beta F^{(0)}}\sum_{n,m} e^{-\beta E_n^{(0)}} |\langle
   \psi_m^{(1)}|\hU(t_f)|\psi_n^{(0)}\rangle|^2
     e^{-\beta (E_m^{(1)}-E_n^{(0)})}\\\nonumber
 &=e^{\beta F^{(0)}}\sum_{n,m} \;
     \langle \psi_m^{(1)}|\hU(t_f)|\psi_n^{(0)}\rangle
     \langle \psi_n^{(0)}|\hUd(t_f)|\psi_m^{(1)}\rangle \; 
     e^{-\beta E_m^{(1)}}\\
 &=e^{\beta F^{(0)}}\sum_m  e^{-\beta E_m^{(1)}}=
   e^{-\beta\DF}\label{wrong}\;.
\end{align}
Hence for this definition of work the Jarzynski equation can easily be
extended to quantum systems. The second law in the form 
$\langle \DE\rangle\ge \DF$ then follows in the usual way from Jensen's
inequality. 

In this approach no genuine quantum observable of work associated with
a self-adjoint operator is introduced. Instead a classical energy
difference is combined with quantum-mechanical transition
probabilities. This makes the proof of the Jarzynski equation very
similar to the case of a classical stochastic process described by a
Master equation \cite{Jar3}. In fact all that is needed from quantum
mechanics is $\sum_n W_{mn}=1$ following from the unitary time
evolution (\ref{eq:defU}).

\section{Operator of work}
In the present setup the work performed is equal to the energy
difference $\DE$ of the system. Since the (time-dependent) energy of a
quantum mechanical system is associated with the Hamilton operator in 
the Heisenberg representation it is reasonable to represent $\DE$ by 
the Hermitian operator \cite{BoKu,Yukawa,AlNi,AlNi2}
\begin{equation}
  \label{eq:defhatW}
  \hDE=\hUd (t_f)\hh(t_f) \hU(t_f) -\hh_0\; .
\end{equation}
The relevant average of a function $f(\DE)$ is then given by  
\begin{equation}
  \label{eq:qmav}
  \langle f(\DE)\rangle=\mathrm{Tr}\, \Big(\hrho\,
         f(\hUd (t_f)\hh(t_f) \hU(t_f) -\hh_0)\Big)\; ,
\end{equation}
where $\hrho$ denotes the equilibrium density matrix corresponding to
$\hh_0$ and Tr stands for the trace. Eq.(\ref{eq:qmav}) represents a
genuine quantum average over the initial condition of the
process. It characterizes the time evolution of the energy of the
system between $t=0$ and $t=t_f$ and does not include contributions
resulting from measurements. 

It is easily shown that the two averages defined in (\ref{eq:clav})
and (\ref{eq:qmav}) respectively give identical results for
functions $f(x)$ involving only linear and quadratic terms. Therefore
both procedures yield the same result for $\langle \DE \rangle$ and are
hence both consistent with the second law. However, whenever $f(x)$
contains higher order non-linearities, as for $f(x)=e^{-x}$, the
results are only the same if the Hamiltonians corresponding to
different times $t$ and $t'$ commute with each other,
$[\hh(t),\hh(t')]=0$. In this case we get indeed  
\begin{align}\nonumber
  \langle e^{-\beta\DE}\rangle
    &=\mathrm{Tr}\,\Big(\hrho\,
      e^{-\beta(\hUd (t_f)\hh(t_f) \hU(t_f) -\hh_0)}\Big)\\\nonumber
    &=e^{\beta F^{(0)}} \mathrm{Tr}\,\Big( e^{-\beta \hh_0}\,
      e^{-\beta(\hUd (t_f)\hh_1 \hU(t_f))}\;
      e^{\beta \hh_0}\Big)\\\nonumber
    &=e^{\beta F^{(0)}} \mathrm{Tr}\,\Big(
      \hUd (t_f)\, e^{-\beta\hh_1}\,\hU(t_f)\Big)= e^{-\beta\DF}
\end{align}
and the Jarzynski equation again holds. However, in general
$[\hh(t),\hh(t')]\neq 0$, therefore 
\begin{equation}
  e^{-\beta(\hUd (t_f)\hh_1 \hU(t_f) -\hh_0)}\neq 
  e^{-\beta\hUd (t_f)\hh_1\hU(t_f)}\,e^{\beta\hh_0}
\end{equation}
and the Jarzynski equation will not be fulfilled. 

We note that the proof of a quantum analogue of the Jarzynski equation
given in \cite{Yukawa} starts out with a definition of a work operator
similar to (\ref{eq:defhatW}). In the subsequent proof, however,
an operator of exponentiated infinitesimal work is introduced which is
somewhat ambigous. It is argued that this ambiguity becomes irrelevant
in the limit $t_f\to\infty$. For general $t_f$, and in particular in
the limit $t_f\to 0$ we are interested in here (see below), however,
this ambiguity disappears only if $[\hh(t),\hh(t+dt)]=0$. 

\section{Exactly solvable limit case}
For the toy model introduced above the implications of the second
definition of work can be analyzed exactly in the limit $t_f\to
0$. The corresponding limit case of the Jarzynski equation is also
known as thermodynamic perturbation \cite{Zwanzig}. In this case of an
{\it instantaneous switch} from $\hh_0$ to $\hh_1$ we have  
\begin{equation}
  \label{eq:limitU}
  \hU(t_f)=1-\frac{i}{\hbar}(\hh_0+\hh_1)t_f+{\cal O}(t^2)\;\to\; 1
\end{equation}
and therefore 
\begin{equation}\label{eq:defDEexample}
  \hat{\DE}=\hh_1-\hh_0=\frac{\om_1^2-\om_0^2}{2}\, \hat{x}^2 \; .
\end{equation}
The work operator is hence diagonal in the position respresentation
and using  for the density operator \cite{LL} 
\begin{align}\nonumber
  \hr_0&(x,x')=\sqrt{\frac{\om_0}{\pi}\tanh\frac{\beta\om_0}{2}}\\
   &\exp{\Big(-\frac{\om_0}{4}\big((x+x')^2\tanh\frac{\beta\om_0}{2}
              -(x-x')^2\coth\frac{\beta\om_0}{2}\big)\Big)}\,  
    \label{eq:rhox}
\end{align}
a simple calculation gives for the Jarzynski estimate $\DJa$ of the
free energy difference
\begin{align}\nonumber
 \DJa&:=-\frac{1}{\beta}\ln \langle e^{-\beta\DE}\rangle\\\nonumber
    &=-\frac{1}{\beta}\ln \int dx \, \hr_0(x,x) \;
     e^{-\frac{\beta}{2}(\om_1^2-\om_0^2)\, x^2}\\
  &=\frac{1}{2\beta}\ln
   \big(1+\beta\,\frac{\om_1^2-\om_0^2}{2\om_0}
               \coth\frac{\beta\om_0}{2}\Big)\, . \label{eq:DJ}
\end{align}
The average of $\DE$ itself is given by  
\begin{equation}
  \label{eq:DW}
  \langle \DE\rangle=
   \frac{\om_1^2-\om_0^2}{4\om_0} \coth\frac{\beta\om_0}{2}\; .
\end{equation}
Comparison of (\ref{eq:DJ}) and (\ref{eq:DW}) shows that for the
present system we have the exact relation 
\begin{equation}
  \label{eq:relation}
  \DJa=\frac{1}{2\beta}\ln(1+2\beta\langle \DE \rangle) \; .
\end{equation}

From (\ref{eq:DF}) and (\ref{eq:DW}) we infer
$\langle\DE\rangle\geq\DF$ for all $\beta$ with the equality holding
only for $\om_1=\om_0$. This is in accordance with the second law of
thermodynamics and the fact that the instantaneous change of $\om$ is
a non-equilibrium process. The inequality holds down to $T=0$ where we
have  
\begin{align}\nonumber
  \langle\DE\rangle&=\langle \psi_0^{(0)}|\hh_1|\psi_0^{(0)}\rangle -
      \langle \psi_0^{(0)}|\hh_0|\psi_0^{(0)}\rangle \\\nonumber
     &\geq
      \langle \psi_0^{(1)}|\hh_1|\psi_0^{(1)}\rangle -
      \langle \psi_0^{(0)}|\hh_0|\psi_0^{(0)}\rangle=\DF
\end{align}
with the inequality sign resulting from the Rayleigh-Ritz variational
principle. 

On the other hand the results (\ref{eq:DF}) and (\ref{eq:DJ}) differ
from each other for all values of $\beta$. Therefore the Jarzynski
equation does not hold for the system under consideration if the
second definition of work is implemented. In view of the discussion in
the previous section this is in accordance with $[\hh_1,\hh_0]\neq 0$.

In the high-temperature limit, $\beta\to 0$, we find
\begin{align}
  \DF&=\frac{1}{\beta}\Big[\ln\frac{\om_1}{\om_0}
       +\frac{\beta^2}{24}(\om_1^2-\om_0^2)+{\cal O}(\beta^4)\Big]\\
  \DJa&=\frac{1}{\beta}\Big[\ln\frac{\om_1}{\om_0}
       +\frac{\beta^2}{24} (\om_1^2-\om_0^2)\frac{\om_0^2}{\om_1^2}
          +{\cal O}(\beta^4)\Big]\; ,
\end{align}
such that the leading behaviour is the same in accordance with the
validity of the Jarzynski equation for classical systems. The first 
differences occur already in the ${\cal O}(\beta^2)$ corrections. Since 
the expansion in $\beta$ is equivalent to an expansion in $\hbar$ we
find that in the classical limit deviations from the Jarzynski
equation occur at order ${\cal O}(\hbar^2)$. A similar result was
obtained recently for quantum corrections to the transient fluctuation
theorem \cite{MoTa}. 

\begin{figure}[t]
\begin{center}
\includegraphics[width=0.45\textwidth]{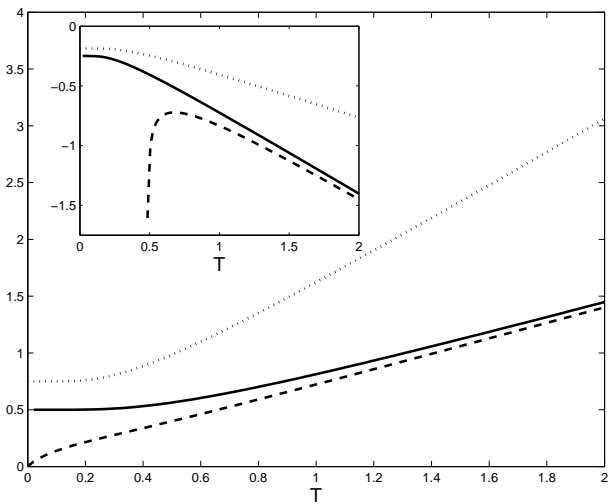}
\end{center}
\caption{\label{fig1} Free energy difference $\DF$ as given by
  (\ref{eq:DF}) (full line), Jarzynski estimate $\DJa$ defined in
  (\ref{eq:DJ}) (dashed), and average work $\langle\delh\rangle_0$
  given by (\ref{eq:DW}) (dotted) as function of the temperature
  $T=1/\beta$ for $\omega_0=1$ and $\om_1=2$. The inset shows the same
  quantities for $\om_0=1$ and $\om_1=0.5$.}
\end{figure}

In the opposite limit, $\beta\to\infty$, we have
\begin{align}
  \DF&\sim \frac{\om_1-\om_0}{2}+\frac{1}{\beta}
         (e^{-\beta\om_0}-e^{-\beta\om_1})\\
  \DJa&\sim \frac{1}{2\beta}
     \ln\Big(1+\beta\frac{\om_1^2-\om_0^2}{2\om_0}\Big)\; ,
\end{align}
showing qualitative differences between the two quantities in the
quantum regime. Whereas $\DF$ tends, as it should, to the difference
between the respective ground-state energies of $\hh_1$ and $\hh_0$ the
estimator $\DJa$ approaches zero (as long as  $\om_1>\om_0$). The
situation is illustrated in Fig.~\ref{fig1}. 

The differences between the two definitions of work for the
system under consideration become particularly clear when considering
the case $\om_1<\om_0$, as illustrated by the inset of
Fig.~\ref{fig1}. Indeed, from (\ref{eq:DJ}) we infer that $\DJa$
diverges at $\beta=\beta_c$ where $\beta_c$ is the solution of
the equation 
\begin{equation}
  \label{eq:defbetac}
  \frac{1}{\beta_c} \tanh(\frac{\beta_c\,\om_0}{2})=
    \frac{\om_0^2-\om_1^2}{2\om_0} \; .
\end{equation}
The origin of this divergence is easily understood
intuitively. For $\om_1<\om_0$ the system gains energy by the change
in $\om$ and accordingly $\hat{\DE}$ as defined in
(\ref{eq:defDEexample}) is negative semi-definite. The system receives
a large amount of energy if 
the particle is far from the origin when the change in $\om$
occurs and these contributions may induce a divergence of 
$\langle e^{-\beta \DE}\rangle$ for sufficiently large $\beta$. 
In the classical case no such divergence may occur since the initial
canonical distribution $\rho_0(x)\sim\exp(-\beta\om_0^2x^2/2)$
confines the particle with increasing $\beta$ to a smaller and smaller
vicinity of the origin which exactly counterbalances the impending
divergence. In the quantum case and using the second definition of
work, however, we find from (\ref{eq:rhox}) in the limit
$\beta\to\infty$ 
\begin{equation}
  \hr_0(x,x) \rightarrow \sqrt{\frac{\om_0}{\pi}}\; e^{-\om_0 x^2}
            =|\psi_0^{(0)}(x)|^2
\end{equation}
and the size of typical excursions of the particle from the origin
becomes independent of $\beta$. Correspondingly, for sufficiently
large values of $\beta$ the Jarzynski average 
$\langle e^{-\beta \DE}\rangle$ will diverge. It is also instructive
to consider why this divergence does not occur in
(\ref{wrong1}). The reason the sum over $n$ cannot diverge is the
same as in the classical case: large values of $E_n^{(0)}$ are
suppressed by correspondingly small values of $P_n^{(0)}$. The $m$-sum
cannot diverge either since for given $n$ the energy difference 
$\DE=E_m^{(1)}-E_n^{(0)}$ becomes positive for sufficiently large
$m$ in spite of $\om_1<\om_0$. In other words, even though the energy
difference is negative semi-definite the measurement of the energy at the
end of the process may yield higher values than the one at the
beginning which renders the expression (\ref{wrong1}) always
convergent. 

In view of the failure of (\ref{eq:JE}) for the second definition of
work it is worthwhile to ask whether there is an operator $\hA$ such
that   
\begin{equation}
  \label{eq:defA}
  \langle e^{-\beta\hA}\rangle=e^{-\beta\DF}\; .
\end{equation}
This operator has to fulfill
\begin{equation}
  \mathrm{Tr}\, e^{-\beta \hA}\,e^{-\beta \hh_0}
     =\mathrm{Tr} \, e^{-\beta \hh_1}
\end{equation}
and one possible choice is
\cite{BoKu,AlNi} 
\begin{equation}
  \label{eq:BHC}
  \hA=-\frac{1}{\beta}\ln \mathrm{T}_>
     \exp\Big(-\int_0^\beta ds \;e^{s\hh_0} (\hh_1-\hh_0)
     \; e^{-s\hh_0}\Big)\; .
\end{equation}
Explicitly we find in the present case 
\begin{equation}
  \label{eq:resA}
  \hat A=\frac{\om_1^2-\om_0^2}{2}\Big[a\,\hat p^2+b\,\hat x^2+
          ic\,(\hat x \hat p +\hat p \hat x)\Big]\; ,
\end{equation}
where the real coefficients $a,b$ and $c$ can be determined from the
Baker-Hausdorff-Campbell \cite{BHC} formula. The first terms are of
the form  
\begin{align}\nonumber
  a=&\frac{\beta^3}{3}+\frac{\beta^5}{30}(\om_0^2+\om_1^2)\\
    &+\frac{\beta^7}{1890}\nonumber
      \Big[11(\om_0^4+\om_1^4)-10\om_0^2\om_1^2\Big]+{\cal O}(\beta^9)\\
  b=&-\beta-\frac{\beta^3}{6}(\om_0^2+\om_1^2)\nonumber
      -\frac{\beta^5}{45}(\om_0^4+\om_1^4+\om_0^2\om_1^2)\\
    &-\frac{\beta^7}{3780}\nonumber
      \Big[13(\om_0^6+\om_1^6)-(\om_0^4\om_1^2 + \om_0^2\om_1^4)\Big]
      +{\cal O}(\beta^9)\\\nonumber
  c=&-\frac{\beta^2}{2}-\frac{\beta^4}{12}(\om_0^2+\om_1^2)
      -\frac{\beta^6}{90}(\om_0^4+\om_1^4)+{\cal O}(\beta^8)\; .
\end{align}
The operator $\hat A$ given by (\ref{eq:resA}) does satisfy
(\ref{eq:defA}) but does not represent a physical observable since
it is not Hermitian. Moreover it is rather specific for the present example.

\section{Conclusion}
The extension of fluctuation theorems or the related non-equilibrium
work relations to quantum systems requires the definition of work done
on the system in a general non-equilibrium process. The aim of the
present paper was to compare the implications of two possible
definitions of work on the generalization of the Jarzynski equation
for a very simple, exactly solvable quantum system. This system is
not in contact with a heat bath during the process and hence the work
performed equals the change in its energy. 

If the system were described by classical mechanics the situation
would be rather clear \cite{note}. The energy difference is given by
the difference between the value of the final hamiltonian at the final 
phase space point and the value of the initial hamiltonian at the initial
phase space point. Since the final state is a unique function of
the initial one the average in the Jarzynski equation is given by the
sole equilibrium average over the initial condition of the system
\cite{Jar}.  

If the system dynamics is described by quantum mechanics at least two
definitions for the work are conceivable. In the first one the the
energy difference of the system is determined via  
measurements at both the beginning and the end of the process
\cite{Kurchan1,Tasaki,Mukamel,TLH}. The energy difference then becomes
a classical stochastic variable with transition probabilities derived
from quantum mechanics and the corresponding average has to combine
the average over the intial condition with the one over the transition
probabilities. In this way the general validity of the Jarzynski
equation can be established similar to the classical case described by
a master equation \cite{Jar3}.  

If on the other hand the energy difference is described by the 
difference between the system Hamiltonian in Heisenberg representation
at the beginning and at the end of the process it represents a genuine
quantum observable \cite{BoKu,Yukawa,AlNi,AlNi2}. In complete analogy
with the classical case the average in the Jarzynski equation is then 
a sole average over the initial condition which is given by the trace
with the initial density operator. This definition of work yields in
the general case results different from the first prescription. In
particular we have shown that for a very simple, exactly solvable
example the Jarzynski identity is violated if the second definition of
work is used. 

The two definitions of work give identical results in the
classical limit $\beta\hbar\to 0$ as well as in the case where initial
and final Hamiltonians commute with each other \cite{JaWo,Kurchan1}.  

At present it seems that the definition of work performed on a quantum
system in a non-equilibrium process is to a certain extend a matter of
taste. We therefore conclude that the question whether
exact and general versions of non-equilibrium fluctuation theorems and
of the Jarzynski equation applicable to quantum systems exist remains
an open and challenging problem. \\

{\bf Acknowledgment:} We are indebted to Chris van den Broeck, Martin
Holthaus, Shaul Mukamel, Theo Nieuwenhuizen, Peter Reimann, and Udo
Seifert for interesting discussions. We would also like to thank the
anonymous referees for stimulating remarks.


\begin{thebibliography}{99}
\bibitem{note} There are some ambiguities in the definition of work
  even at the classical level, see \cite{Jar4} for a clear discussion,
  which, however, do not bear on the problem discussed in the
  present paper.
\end{thebibliography}
\end{document}